\documentstyle[12pt,psfig]{article}

\begin{document}
\vskip 1.5cm
\centerline{\large\bf Gluon- vs. Sea quark-Shadowing}
\medskip
\bigskip
\medskip
\centerline{\bf N. Hammon, H. St\"ocker, W. Greiner \footnote{Work supported
by BMBF, DFG, GSI}}
\medskip
\bigskip
\centerline{Institut F\"ur Theoretische Physik}
\centerline{Robert-Mayer Str. 10}
\centerline{Johann Wolfgang Goethe-Universit\"at}
\centerline{60054 Frankfurt am Main}
\centerline{Germany}
\medskip
\bigskip
\medskip
\bigskip
%\centerline{\Large\bf -- DRAFT -- }
%\centerline{\today}
\bigskip
\centerline{Abstract}
\bigskip
{\small
We calculate the shadowing of sea quarks and gluons and show that the 
shadowing of gluons is {\it not} simply given by the sea quark shadowing,
especially at small $x$.
The calculations are done in the lab frame approach by using the generalized vector meson
dominance model. Here the virtual photon turns into a hadronic fluctuation long before
the nucleus. The subsequent coherent interaction with more than one nucleon in the nucleus
leads to the depletion $\sigma (\gamma^* A) < A\sigma (\gamma^* N)$ known as shadowing.
A comparison of the shadowing of quarks to E665 data for $^{40}Ca$ and 
$^{207}Pb$ shows good agreement.}
\newpage
\centerline{\bf 1. Introduction}

When calculating perturbative QCD cross sections in nucleus nucleus collisions
one has to take care of an additional effect not appearing on the pure
nucleon nucleon level: nuclear shadowing. As a result of this depletion
of the nuclear parton densities at small $x$ one finds a strong suppression 
of e.g. charmonium states and minijets \cite{nils1,kari1}. Also for prompt
photons smaller multiplicities result \cite{nils2} due to the smaller number density
of partons in the relevant region of the momentum fraction variable.\\
In either model
(lab frame vector meson dominance type models or infinite momentum frame parton fusion 
models) the shadowing of gluons is expected to be much stronger than the shadowing of sea 
quarks, even at small $x$. This seems to contradict the first naive expectation in terms 
of the
QCD improved parton model that explains the scaling violation of the structure functions
via the DGLAP splitting functions which treat sea quarks similar to gluons
(in the sense that the sea quarks are produced by the gluons when $Q^2$ increases)
in a region where essentially no momentum is carried by
valence quarks, i.e.~in the small $x$ region.\\
In the following we will focus on the lab frame interpretation which explains
the shadowing phenomenon by use of the generalized vector meson dominance model (GVMD).\\
\centerline{\bf 2. Lab frame description of shadowing}

In the lab frame description one essentially makes use of the hadronic structure
of the virtual photon, manifesting itself in a field theoretic approach in two different 
time orderings assuring gauge- and Lorentz invariance to the amplitudes \cite{piller}.
At small enough $x$ ($x\ll 0.1$) the hand-bag graph contribution gets small.
The interaction then proceeds via the VMD graph where the virtual
photon fluctuates into a $q\bar q$ pair within the "coherence time" 
$\lambda \approx 2\nu /(\mu^2 +Q^2) \approx 1/(2mx)$, where $m$ is the nucleon mass,
$\mu$ the mass of the pair and $\nu$ the lepton energy loss. At small $Q^2$ the
interaction with a free nucleon then proceeds via the $\rho , \omega$, and $\phi$ mesons.
At larger $Q^2$ the $q\bar q$ continuum has to be taken into account to describe the $F^{N}_{2} (x,Q^2)$
data. Including sea quarks $xf_N (x,Q^2)$ only, the nucleon structure function can 
(for small $x$) be written as \cite{ina}
\begin{equation}
F_{2}^{N} (x,Q^2) = \sum _{i} e_{i}^{2} xf_N (x,Q^2)=\frac{Q^2 \sigma (\gamma ^{*}N)}
{4\pi ^2 \alpha _{em}}
\end{equation}
With the transverse size of the $q\bar q$ pair frozen during the scattering, 
due to Lorentz contraction, the 
photon-nucleon cross section can be factorized:
\begin{equation}
\sigma (\gamma ^{*}N) = \int_{0}^{1}dz \int d^2 r \left |\psi (z,r)\right|^2 
\sigma_{q\bar qN}(r)
\end{equation}
with the Sudakov variable $z$. The photon wave function $\left |\psi (z,r)\right|^2$ 
can be interpreted as 
the propability of finding a $q\bar q$ state with transverse separation $r$ and a 
momentum fraction $z$ and $(1-z)$ with respect to the virtual photon. The wave function 
can be expressed as \cite{ina} 
\begin{equation}
\left |\psi (z,r)\right|^2 = \frac{3}{2}\frac{\alpha_{em}}{\pi ^2}\sum _{i=1}^{N_f}
e_{i}^2 P(z) a^2 K_{1}(ar)^2
\end{equation}
where $P(z)=z^2+(1-z)^2$ is the splitting function of the gauge boson into the $q\bar q$
pair, $a^2=Q^2 z(1-z)$ and $K_1$ is the modified Bessel function. 
The most important contributions come from the region where
$ar$ is small. Therefore pairs with a small transverse size are favored.\\
The cross section for the strong interaction of the pair with the nucleon,
which happens via gluon exchange, is \cite{lonya}
\begin{equation}
\sigma_{q\bar qN} = \frac{\pi^2}{3}r^2 \alpha_s (Q'^2)x'g_N (x',Q'^2)
\end{equation}
with $x' = x/a^2r^2$ and $Q'^2 = 4/r^2$. The strong scaling violation of $x'g_N (x',Q'^2)$
at small $r$ (large $Q'^2$) in turn compensates for the smallness of the pair
(for further details see \cite{ina} and references therein). 
Therefore the exact expression for the sea quark density is
\begin{equation}
x f_N (x,Q^2) = \frac{3}{4\pi^3}x \int_{x}^{1} \frac{dx'}{x'^2}\int_{4/Q^2}^{\infty} 
\frac{dr^2}{r^4} \sigma_{q\bar qN} (x',r^2)
\end{equation}
The main problem at this point arises from the interplay between hard and soft physics.
In some region $4/Q^2 \le r^2 \le 4/Q_{0}^2$ perturbation theory should be valid. For 
$r^2 > 4/Q_{0}^2$, i.e.~for small $Q^2$, the non-perturbative part dominates.
To use the presently available parametrizations of the free nucleon parton densities
we choose a lower cut-off of $Q_{0}^2 = 0.4$ GeV$^2$. It is obvious that we thereby 
completely neglect the non-perturbative input below this scale. However, 
in \cite{piller} it was shown
that for $Q^2 \leq 10^{-1}$ GeV$^2$ a saturation in the shadowing ratio
sets in. As a result, the uncertainty by choosing a cut-off at $Q_{0}^2 = 0.4$ GeV$^2$ is
on the few percent level. Also, when comparing to 
experimental results, we find good agreement with the data.\\
In the case of the gluon distribution of the free nucleon one considers the scattering
of a virtual colorless probe (e.g.~the virtual higgs boson) which proceeds via the 
production of a gluon pair which then strongly interacts with the nucleon. 
With the momentum cut-off the gluon density can be written as 
\begin{equation}
x g_N (x,Q^2) = \frac{4}{\pi^3} \int_{x}^{1} \frac{dx'}{x'} \int_{4/Q^2}^{4/Q_{0}^{2}} 
\frac{dr^2}{r^4} \sigma_{gg N} (x',r^2)
\end{equation}
The main difference between the scattering of the $q\bar q$ pair and the $gg$ pair stems
from the much larger cross section $\sigma_{gg N} = 9/4 ~\sigma_{q\bar qN}$ which in turn 
leads to a much stronger gluon shadowing.\\

In the case of deep inelastic scattering off nuclei the same relations 
hold but with the respective cross sections for the scattering of the hadronic and 
gluonic fluctuations off the nucleus. For the sea quark distribution in the nucleus $A$ 
one has
\begin{equation}
x f_A (x,Q^2) = \frac{3}{4\pi^3}x \int_{x}^{1} \frac{dx'}{x'^2}\int_{4/Q^{2}}^{4/Q_{0}^{2}}
\frac{dr^2}{r^4} \sigma_{q\bar qA} (x',r^2)
\end{equation}
and for the gluon the resulting equation is given by
\begin{equation}
x g_A (x,Q^2) = \frac{4}{\pi^3} \int_{x}^{1} \frac{dx'}{x'} \int_{4/Q^{2}}^{4/Q_{0}^{2}}
\frac{dr^2}{r^4} \sigma_{gg A} (x',r^2)
\end{equation}
As stated above, it is the long distance phenomenon of the hadronic fluctuation which causes 
the shadowing effect. For the small $x$, i.e.~large $\lambda$, the fluctuation interacts
with more than one nucleon inside the nucleus. As a result of this coherent interaction
one finds that $\sigma (\gamma^* A) < A \sigma (\gamma^* N)$.
The specific feature that is responsible for the shadowing effect is the fact that
$\sigma_{hA} \neq A \sigma_{hN}$. Within Glauber theory in the eikonal limit one 
finds \cite{bauer}
\begin{equation}
\sigma_{hA} = 2 \int d^2b \left( 1-e^{-\sigma_{hN} T_A (b)/2} \right)
\end{equation}
with the nuclear thickness function
\begin{equation}
T_A (b) = \int dz \rho _A (b,z) = \frac{A}{\pi R_{A}^2}e^{-b^2 / R_{A}^2}
\end{equation}
The integration can be done exactly for a Gaussian parametrization
of $T_A (b)$. Here we will include only the double scattering
contribution which was shown \cite{ina} to strongly dominate the overall shadowing ratio
$R_G = g_N (x,Q^2) / g_A$; the triple and higher scattering terms give a contribution
that is only on the percent level and is therefore neglected here.
We use \cite{piller,gribov}
\begin{equation}
\sigma_{hA} = A \sigma_{hN} \left[ 1- A^{1/3} \frac{\sigma_{hN}}{8\pi a^2}\frac{A-1}{A}
{\rm exp} \left(-\frac{a^2 A^{2/3}}{2\lambda ^2}\right)+\dots \right]
\end{equation}
with $a=1.1fm$.\\
\centerline{\bf 3. Results}

\begin{figure}
\centerline{\psfig{figure=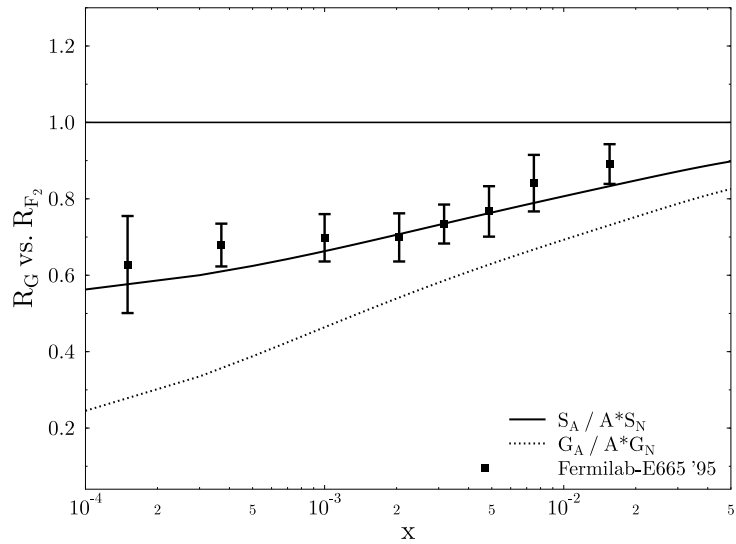,width=14cm}}
\caption{$R_{F_2}$ vs. $R_G$ at $Q^2 = 4$ GeV$^2$ for $^{207}Pb$.}
\label{ratio}
\end {figure}

\begin{figure}
\centerline{\psfig{figure=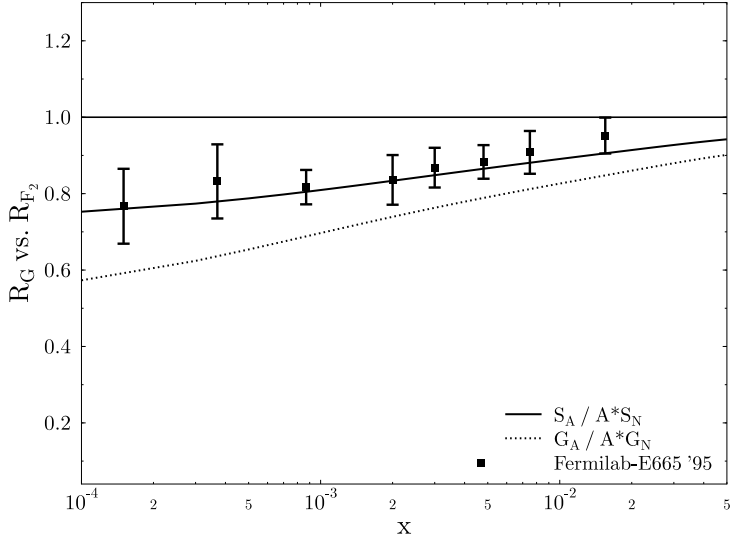,width=14cm}}
\caption{$R_{F_2}$ vs. $R_G$ at $Q^2 = 4$ GeV$^2$ for $^{40}Ca$.}
\label{ratio2}
\end {figure}

\begin{figure}
\centerline{\psfig{figure=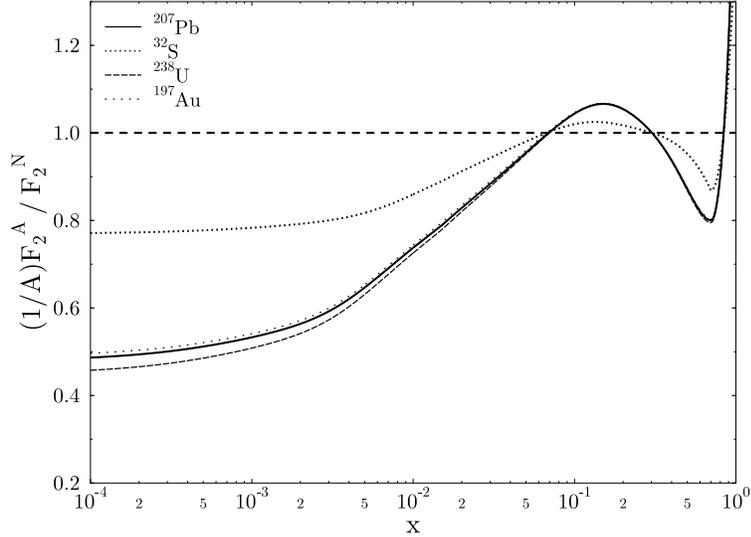,width=14cm}}
\caption{$R_{F_{2}}=F_{2}^{A}/A F_{2}^{N}$ as parametrized by Eskola in 
\protect\cite{kari1} at 
$Q^2=4$ GeV$^2$.}
\label{shadow}
\end {figure}

Based on the approximations above, namely double scattering contribution only, a
lower momentum cut-off at $Q^{2}_{0}=0.4$ GeV$^2$, due to the lack of information on the 
non-perturbative input, and by employing the Gl\"uck, Reya, Vogt parametrization
\cite{grv} we derived the shadowing ratios for quarks and gluons
at the typical semi-hard scale $Q=2$ GeV for $^{207}Pb$ and
$^{40}Ca$ (see figures \ref{ratio} and \ref{ratio2}). 
One clearly sees that gluons are much stronger shadowed 
than sea quarks at small $x$ which is due to the effects outlined above. 
This is a very important feature since at the soon
available collider energies one will particularly test the small $x$ region 
($x=p_T /\sqrt{s}$ at midrapidity) which
becomes increasingly important due to the strong gluonic component inside the free
nucleon and therefore in the heavy nuclei.\\
This feature has a large impact on transverse energy production and on 
the initial temperature and entropy produced in heavy ion collisions. It should therefore 
come as no surprise when the final particle 
multiplicities are much lower than expected from calculations using e.g.~the quark 
shadowing ratio $R_{F_2} (x)$ shown in figure \ref{shadow} for gluons also.
To check our results, we compared the results for $R_{F_2} (x)$ to Fermilab-E665 
data \cite{fermilab} and find a good agreement. Since $s\simeq Q^2/x$, for deep inelastic
events, one is
restricted to a certain $Q^2$-range for the different $x$-values in the experiment.
To account for this feature we compared to the data within the range
$\left< Q^2\right> \approx 2.42$GeV$^2$ - 4.45GeV$^2$ because our calculation was done
at fixed $Q^2 = 4$ GeV$^2$. One also sees that the agreement with the data is only good 
at small enough $x$, i.e.~at large enough coherence lengths $\lambda \sim 1/\Delta k_z$, where
$\Delta k_z = k^{\gamma}_{z}-k^{h}_{z}$ is the phase shift between the photon and the 
hadronic fluctuation. 
To derive the shadowing ratio for the complete structure function $F_2$
one could in principle improve the large $x$ behavior by
inserting a splitting function $P_{gq} = z^2+(1-z)^2$ as suggested in \cite{ina} where 
$z=x/x'$. Also one has to include the 
valence quark distribution as $F_{2}^N (x,Q^2) = \sum_{i} e_{i}^2 \left[ xf_N (x,Q^2)
+xv_i (x,Q^2)\right]$. But we here neglect this point since we were mainly interested
in the region of small $x$.
\bigskip

\centerline{\bf 4. Conclusions}
We calculated the shadowing of 
sea quarks and gluons at small values of $x$ in the generalized vector meson 
dominance model (GVMD) approach. Based on the Glauber-Gribov multiple scattering
picture for a $q\bar q$ pair with a coherence length
$1/ \Delta k_z = \lambda \approx 1/2mx$, we showed that one should expect significant
differences between $R_{F_2}$ and $R_G$ at small $x$. 
This feature, often neglected in the past, has severe consequences on quantities such as
charmonium production, minijets and therefore on the production of transverse energy and 
entropy. When applying the much stronger gluon shadowing one should expect significantly
smaller multiplicities for processes involving gluons in the initial state compared to 
calculations that accounted for all the shadowing effects by simply employing the quark
shadowing ratio $R_{F_2}=F_{2}^{A}/A F_{2}^{N}$ for all parton species.\\

\end{document}